\documentclass[conference]{IEEEtran}
\IEEEoverridecommandlockouts

\def\IEEEtitletopspace{18pt}
\usepackage{cite}
\usepackage{amsmath,amssymb,amsfonts}
\usepackage[margin=0.75in]{geometry}
\usepackage{algorithmic}
\usepackage{graphicx}
\usepackage{array}
\usepackage{booktabs}
\usepackage{longtable}
\usepackage{subcaption}
\usepackage{xurl}             
\usepackage[breaklinks]{hyperref}
\usepackage{textcomp}
\usepackage{xcolor}
\usepackage{lipsum}
\usepackage{soul}

\def\BibTeX{{\rm B\kern-.05em{\sc i\kern-.025em b}\kern-.08em
    T\kern-.1667em\lower.7ex\hbox{E}\kern-.125emX}}
\begin{document}

\title{Extracting Insights from Large-Scale Telematics Data for ITS Applications: Lessons and Recommendations\\
}

\author{ Gibran~Ali,
Neal~Feierabend,
Prarthana~Doshi,
Calvin~Winkowski,
Michael~Fontaine
\thanks{G. Ali*, N. Feierabend, P. Doshi, and C. Winkowski are with the Virginia Tech Transportation Institute (*email: gali@vtti.vt.edu)}
\thanks{M. Fontaine is with the Virginia Department of Transportation}
}

\maketitle

\begin{abstract}
Over 90\% of new vehicles in the United States now collect and transmit telematics data. Similar trends are seen in other developed countries. Transportation planners have previously utilized telematics data in various forms, but its current scale offers significant new opportunities in traffic measurement, classification, planning, and control. Despite these opportunities, the enormous volume of data and lack of standardization across manufacturers necessitates a clearer understanding of the data and improved data processing methods for extracting actionable insights.  

This paper takes a step towards addressing these needs through four primary objectives. First, a data processing pipeline was built to efficiently analyze 1.4 billion miles (120 million trips) of telematics data collected in Virginia between August 2021 and August 2022. Second, an open data repository of trip and roadway segment level summaries was created. Third, interactive visualization tools were designed to extract insights from these data about trip-taking behavior and the speed profiles of roadways. Finally, major challenges that were faced during processing this data are summarized and recommendations to overcome them are provided. This work will help manufacturers collecting the data and transportation professionals using the data to develop a better understanding of the possibilities and major pitfalls to avoid. 
\end{abstract}


\section{Introduction}

Intelligent Transportation System (ITS) planners have traditionally relied on data sources such as inductive loops, radar sensors, probe vehicles, and traffic cameras for traffic monitoring, planning, and traffic control. However, over the last few years the proportion of vehicles equipped with telematics data capability has been steadily increasing. Now, over 90\% of new vehicles sold in the US have some form of original equipment manufacturer (OEM)-installed telematics data service \cite{berg2024telematics}. These vehicles record and transmit information about location, speed, acceleration, and various vehicle states every few seconds, which is then processed and stored on data servers. The scale, resolution, and accuracy of this data presents a transformative opportunity for the field to accurately measure the state of traffic using a single source of data and use it for planning and control activities.

Within the next few years, such data could be available for the majority of the vehicles. To effectively operationalize this data at scale, it is important to develop a deeper understanding of the data, as well as likely use cases and potential challenges. 

The North American intelligent traffic management system market, valued at USD 3.73 billion in 2021, is projected to expand at 11.5\% annually through 2030 \cite{grandview2022itms}. Major data aggregators like Wejo, during its operation, processed 17 billion daily data points from 12 million vehicles, achieving nearly 95\% of road coverage \cite{wejo_wired_mobility}. Innovations such as edge computing frameworks now enable real-time applications, including driver behavior profiling and dynamic alerts, surpassing traditional systems’ capabilities \cite{8377916}. Early implementations, including Wejo’s Real-Time Traffic Intelligence and GeoTab’s fleet optimization tools, demonstrated telematics’ potential for congestion detection, safety monitoring, and operational efficiency, leveraging billions of daily data points to refine workflows and urban planning strategies \cite{privacyanalytics_geotab_2022}.


ITS planners and departments of transportation (DOTs) have many potential uses for telematics data \cite{Ghaffarpasand2022, Hassan2025,Gao2022}. These data can be used to understand speed and other kinematic profiles experienced on each roadway segment. Such information would provide a much more granular understanding of speeding behavior on a wide variety of roadways \cite{ghaffarpasand2023telematics}. Telematics data could also be used to understand traffic patterns by hour of day, day of week, and season. Such data could also be utilized to validate existing metrics such as travel time estimates that are often calculated by third-party providers with proprietary methodologies \cite{chowdhury2024data}. 

One of the key challenges in operationalizing this data is building efficient data processing pipelines that can handle data at scale \cite{ma2023overview}. It is essential to understand how  elements of these pipelines will scale as the amount of data going through them increases. Such understanding will be helpful in developing applications that are responsive while handing an ever increasing scale of data. This includes developing an understanding of storage system and computational requirements, processing latency, and ideal schema for various types of queries \cite{chowdhury2024data}. 

Developing an understanding of telematics use cases and data processing pipelines will also be helpful for OEMs as they explore the various monetization strategies for the data. Once consensus develops around these topics, various OEMs could standardize data storage and processing steps so that data from multiple OEMs could be seamlessly integrated. Such standardization can reduce overall costs and protect user privacy without compromising  data usability.

The Virginia Department of Transportation (VDOT) procured 6 months of telematics data from Virginia to better understand the various use cases and data processing needs. The work presented in this paper operationalizes this data and fulfills three goals. First, it develops a deeper understanding of the data and the processing requirements. Second, it successfully tests two use cases utilizing this data: a roadway segment level speed distribution and travel patterns by  zip code and hour of day. Finally, it summarizes key lessons and recommendations for using such data.

\section{Data}

The data were acquired by VDOT from a telematics data aggregator that collects data from multiple OEMs and standardizes the variables. Information about vehicle manufacturer and vehicle type was not available. The data also did not uniquely identify vehicles, and therefore multiple trips could not be connected through a vehicle identifier. The data aggregator was asked to provide any passenger vehicle data that passed through Virginia for select data ranges between August 2021 and August 2022. This included full trips that were either completely or partially collected in the state.

The dataset consisted of two types of data: movement and event. Movement data consisted of a unique trip identifier (\textit{journey\_id}), a UTC timestamp, vehicle heading and speed, latitude and longitude coordinates, and vehicle ignition status. The movement data also consisted of metadata such as geohash, postal code, and country code. The average frequency of this data was one data point every three seconds (0.33 Hz). This data consisted of 59.17 billion data points with 120.96 million \textit{journey\_ids}.

The event data consisted of various types of events ranging from vehicle states such as ignition status, wiper activations, seat belt use and turn signal states, to safety-relevant events such as hard braking and automated emergency braking activations. Each data point consisted of a unique identifier, a \textit{journey\_id}, a timestamp, location, vehicle heading and speed, event-specific metrics, and metadata. There were a total of 1.25 billion such events from 83.16 \textit{journey\_ids}. However, only 7.42 million \textit{journey\_ids} were common between the movement and event datasets.

The process of assigning \textit{journey\_id} for a set of points was described by the data aggregators and therefore could have differed for data from one OEM to another. For the purposes of this paper, the \textit{journey\_ids} are used as they existed in the dataset. However, certain data-cleaning steps were used to remove \textit{journey\_ids} that were unlikely to be actual trips: the trip length should have been longer than 100 meters and the trip durations should have been shorter than 24 hours. This removed about 5 million \textit{journey\_ids}, most of which were one- or two-data point trips.

\begin{figure}[t]
    \centering
    \includegraphics[width=1\columnwidth,keepaspectratio]{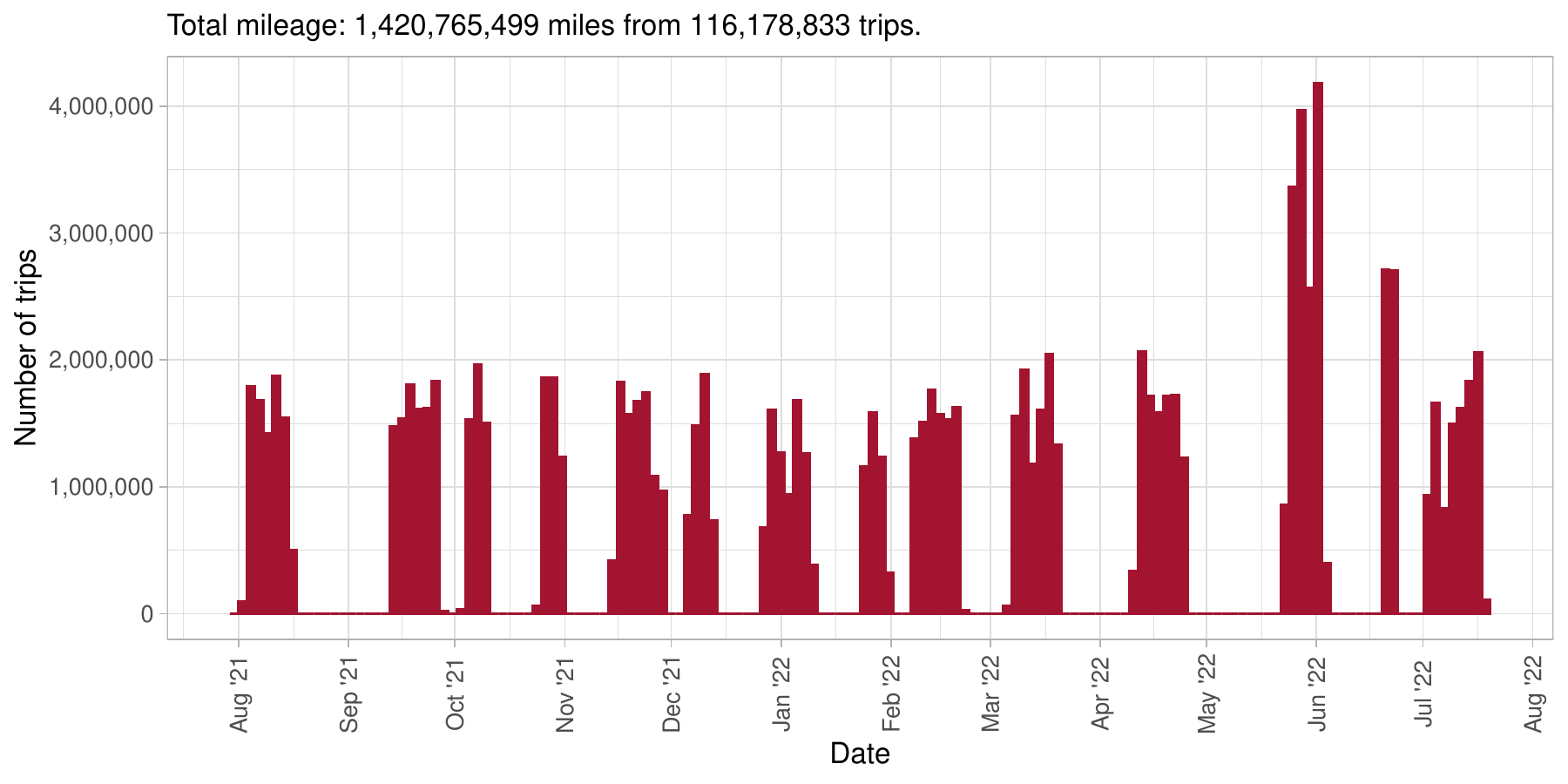}
\caption{Number of trips by date of travel.}
\label{fig:trip-sum-date}
\end{figure}

\begin{figure}[t]
    \centering
    \includegraphics[width=1\columnwidth,keepaspectratio]{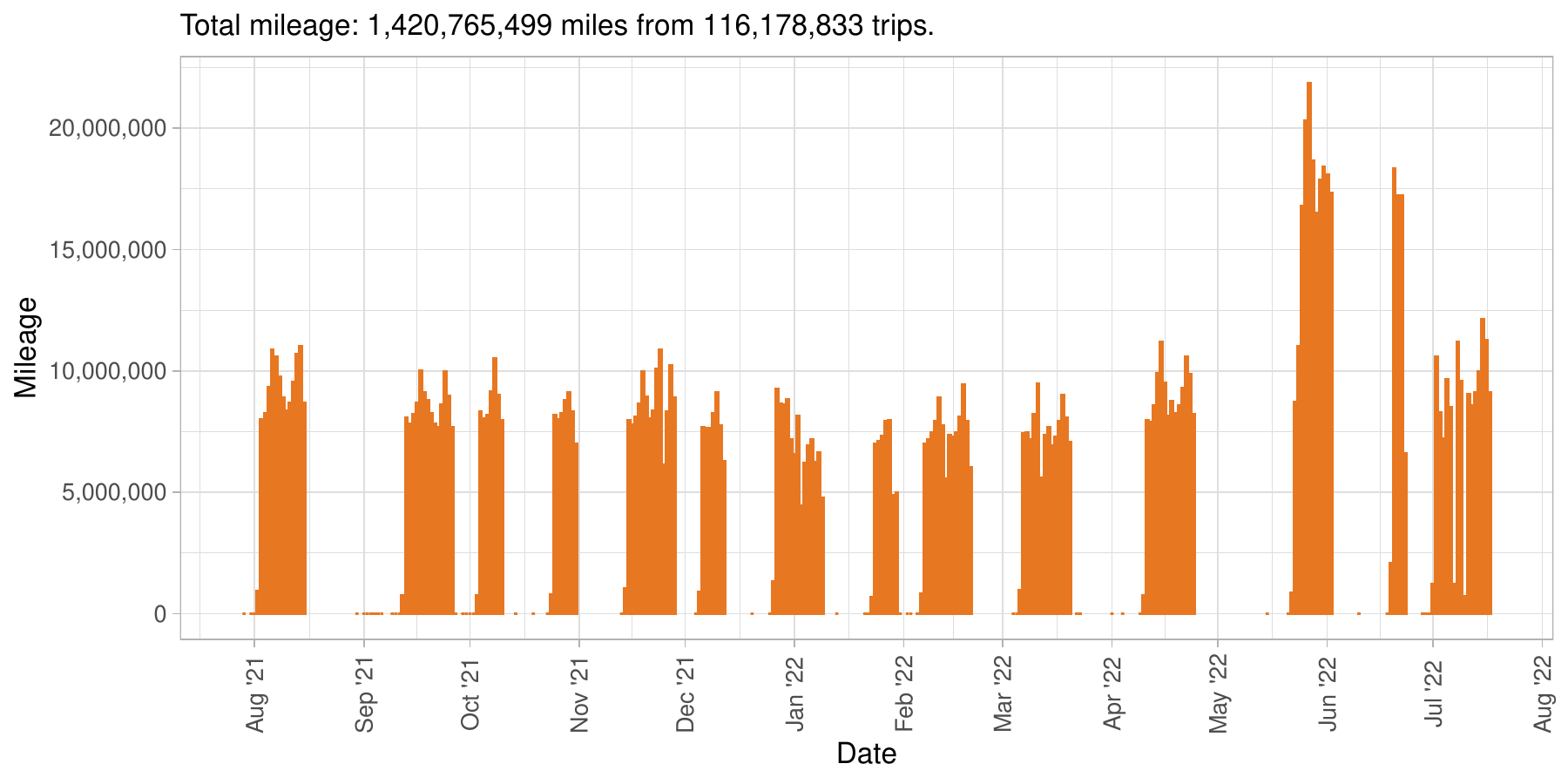}
\caption{Number of miles by date of travel.}
\label{fig:mileage-sum-date}
\end{figure}

Figures \ref{fig:trip-sum-date} and \ref{fig:mileage-sum-date} illustrate the number of trips and mileage collected by date of travel in Virginia. The data represents 24 weeks of driving collected over a year, starting August 2021 to the end of July 2022. The dates were selected to represent driving in various seasons and included dates where major weather events had occurred. The number of trips and mileage accumulation follow similar trends across the date range. A spike in the number of trips and mileage can be seen around June 2022. The reason for this is not fully understood and could be due to an increase in trips or data from additional manufacturers becoming available. Since the data aggregator did not provide this information, the unknown sampling biases remain a major limitation of this dataset. However, due to the large scale of the data, the insights about data processing, trip-taking behavior, and roadway-level summarization are still useful for many ITS and DOT applications.

\begin{figure}[h]
    \centering
    \includegraphics[width=1\columnwidth,keepaspectratio]{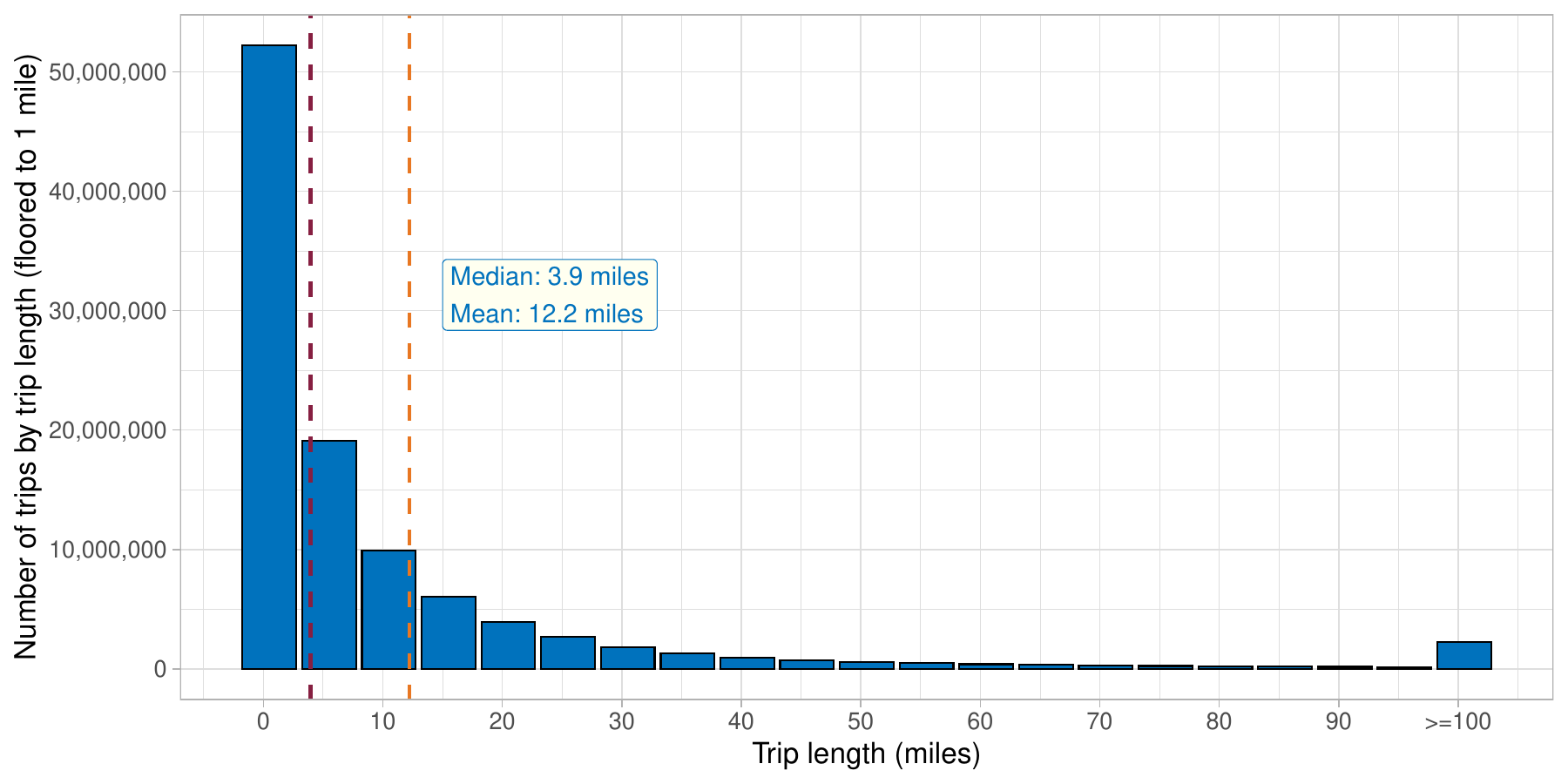}
\caption{Number of trips by trip length with trips longer than 100 miles grouped into a single bin.}
\label{fig:num-trip-length}
\end{figure}

\begin{figure}[h]
    \centering
    \includegraphics[width=1\columnwidth,keepaspectratio]{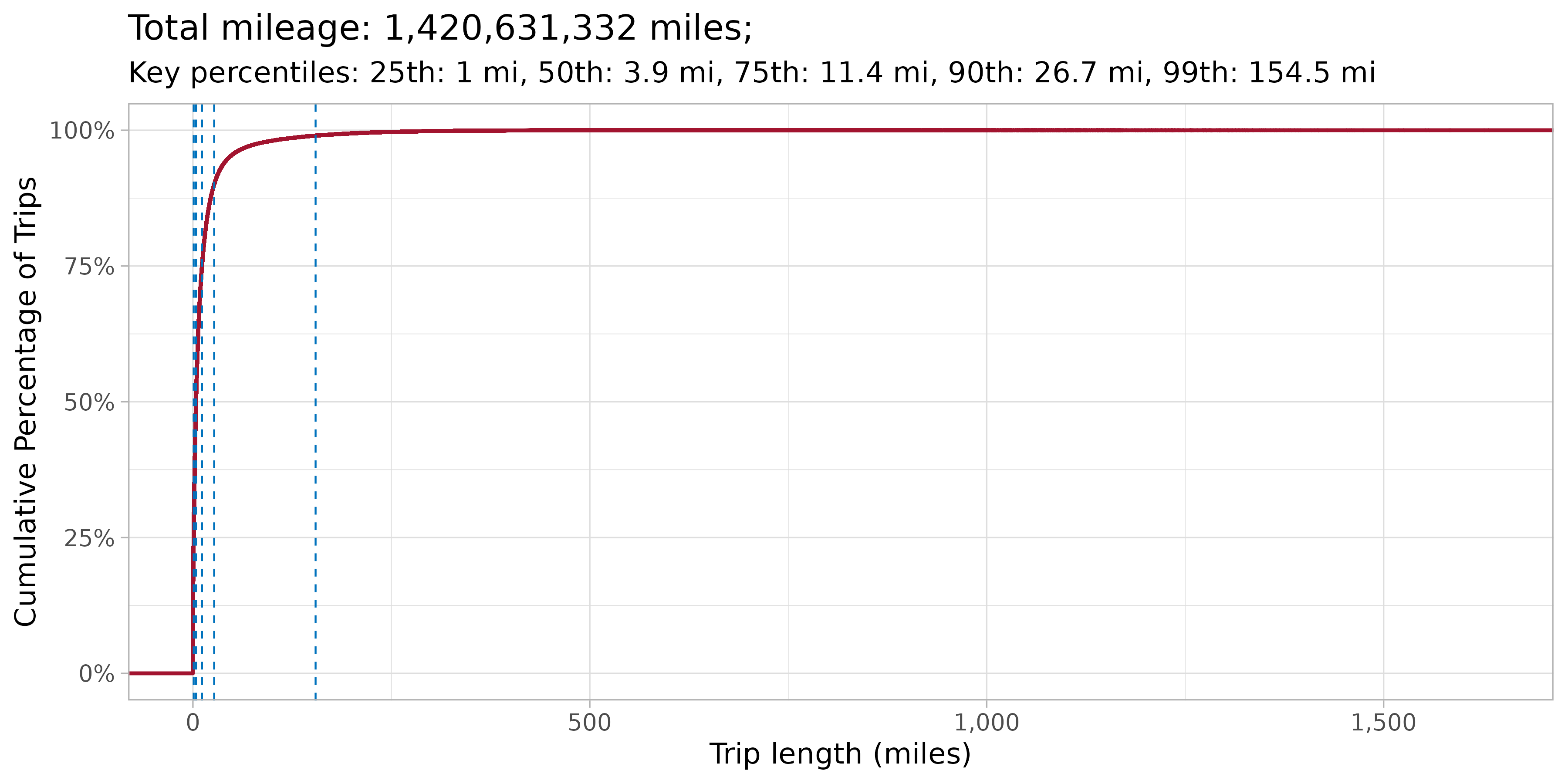}
\caption{Cumulative percentage of trips by trip length.}
\label{fig:num-trip-length-cfd}
\end{figure}

\begin{figure}[h]
    \centering
    \includegraphics[width=1\columnwidth,keepaspectratio]{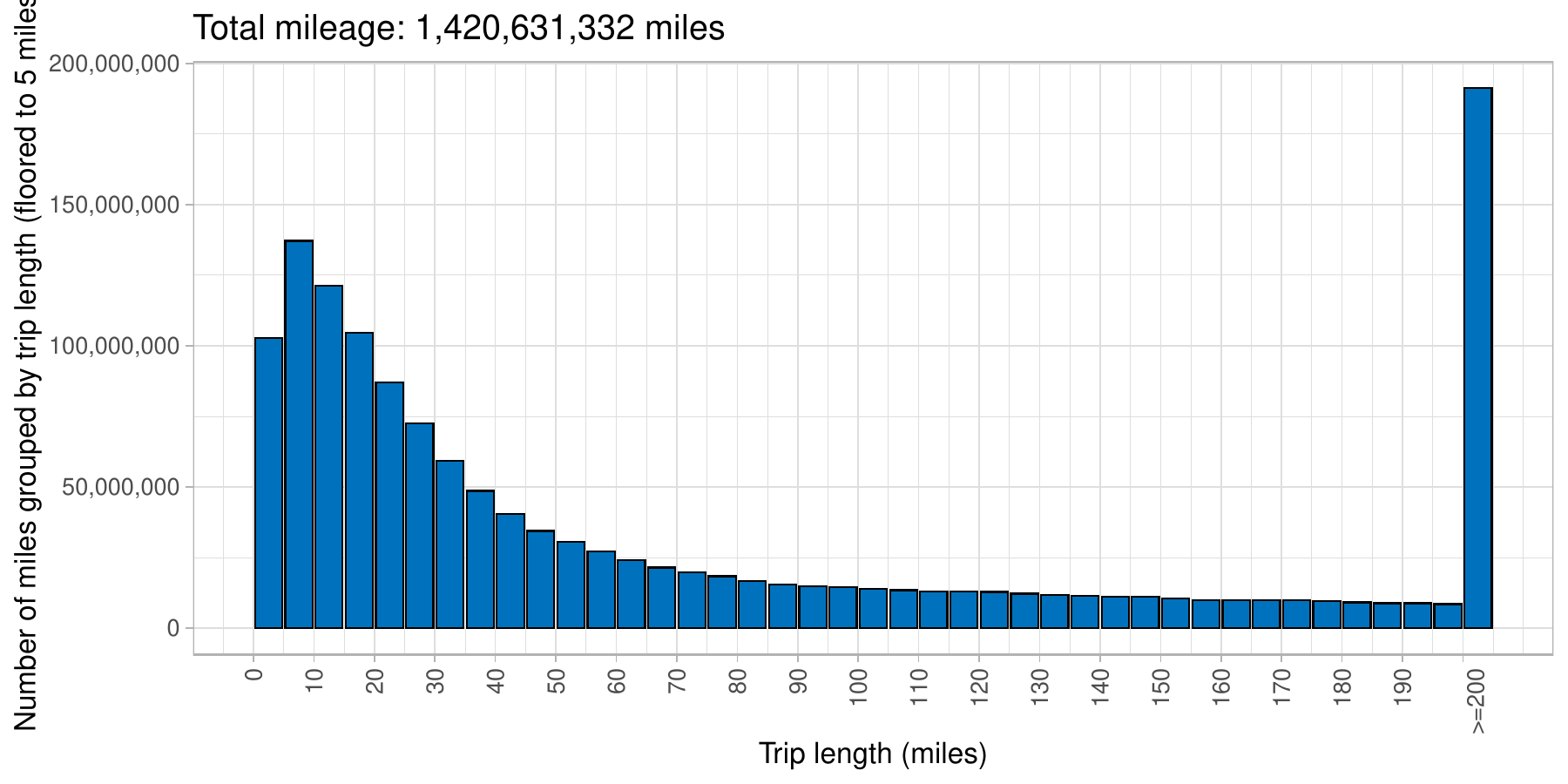}
\caption{Number of miles collected by trip length bin with trips longer than 200 miles grouped into a single bin.}
\label{fig:num-mile-length}
\end{figure}

\begin{figure}[h]
    \centering
    \includegraphics[width=1\columnwidth,keepaspectratio]{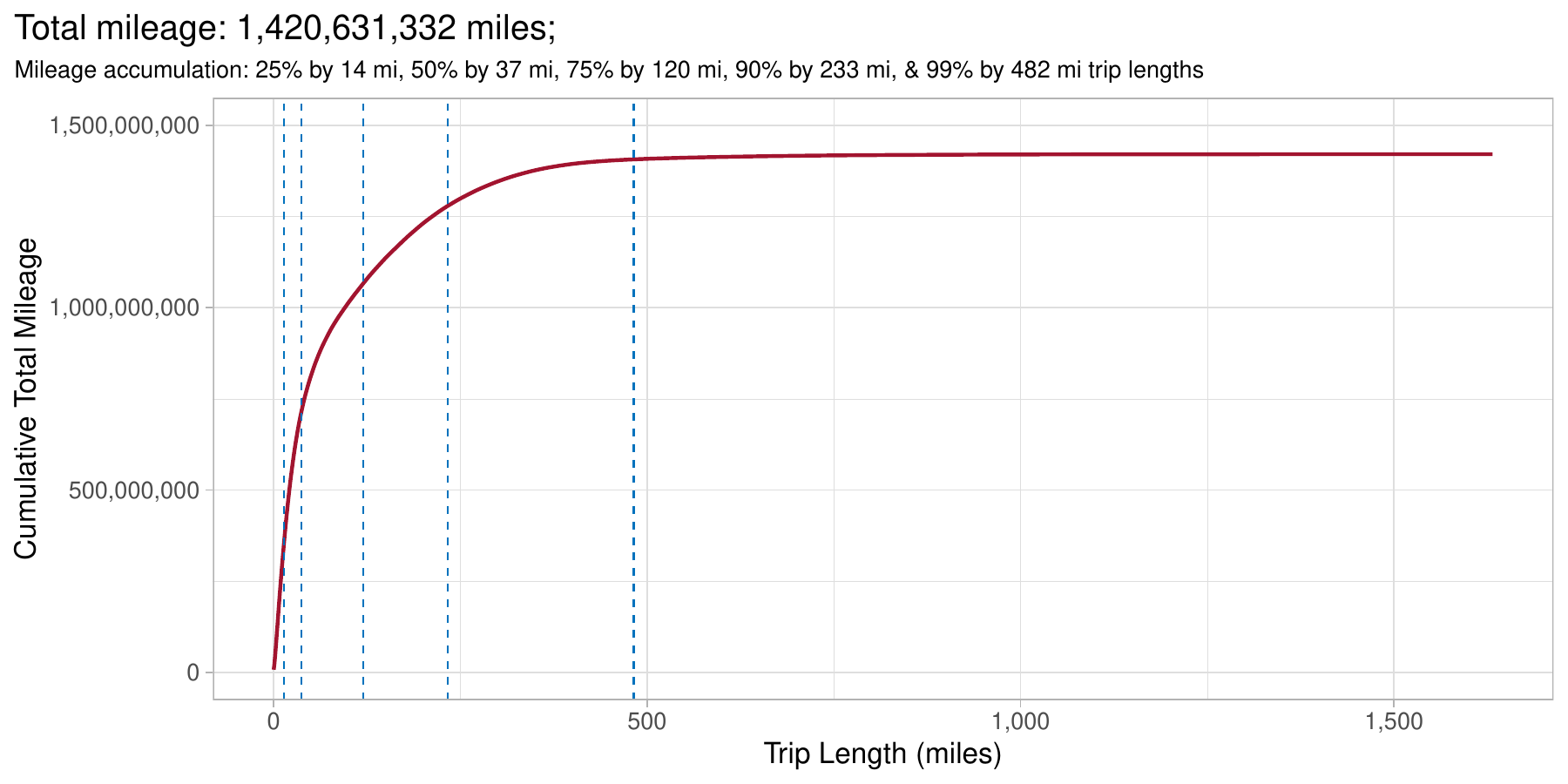}
\caption{Cumulative percentage of miles by trip length.}
\label{fig:num-mile-length-cfd}
\end{figure}

Figure \ref{fig:num-trip-length}  illustrates the number of trips in various mileage bins. For example, about 52 million trips had mileage between 0 and 5 miles. The median trip was 3.3 miles and the 75th-percentile trip was about 12.2 miles. Trips longer than 100 miles were aggregated into a single bin. Figure \ref{fig:num-trip-length-cfd} illustrates similar information in the form of cumulative percentage of trips by mileage. 

Figure \ref{fig:num-mile-length} illustrates the mileage accumulation by trip length bin. For example, trips in the first bin of 0-5 miles account for just over 100 million miles of the data. Trips longer than 200 miles were grouped into one bin and constitute the largest fraction of mileage in the plot. Figure \ref{fig:num-mile-length-cfd} shows  the cumulative percentage of mileage accumulated by different trip lengths. When comparing Figures \ref{fig:num-trip-length} to \ref{fig:num-mile-length-cfd}, it is interesting to see that trips up to the 90\textsuperscript{th} percentile of 26.7 miles constitute less than 50\% of the total mileage. 

To verify the accuracy of the data, trips longer than 1,000 miles were randomly sampled and visualized to validate the data. However, whether these long trips included rest stops or other types of breaks has not been examined. As long as the data was linked together through a single \textit{journey\_id} and had reasonable relationships between duration and distance based on vehicle speed and GPS points, the trip was considered valid for this analysis.

\begin{figure}[t]
    \centering
    \includegraphics[width=1\columnwidth,keepaspectratio]{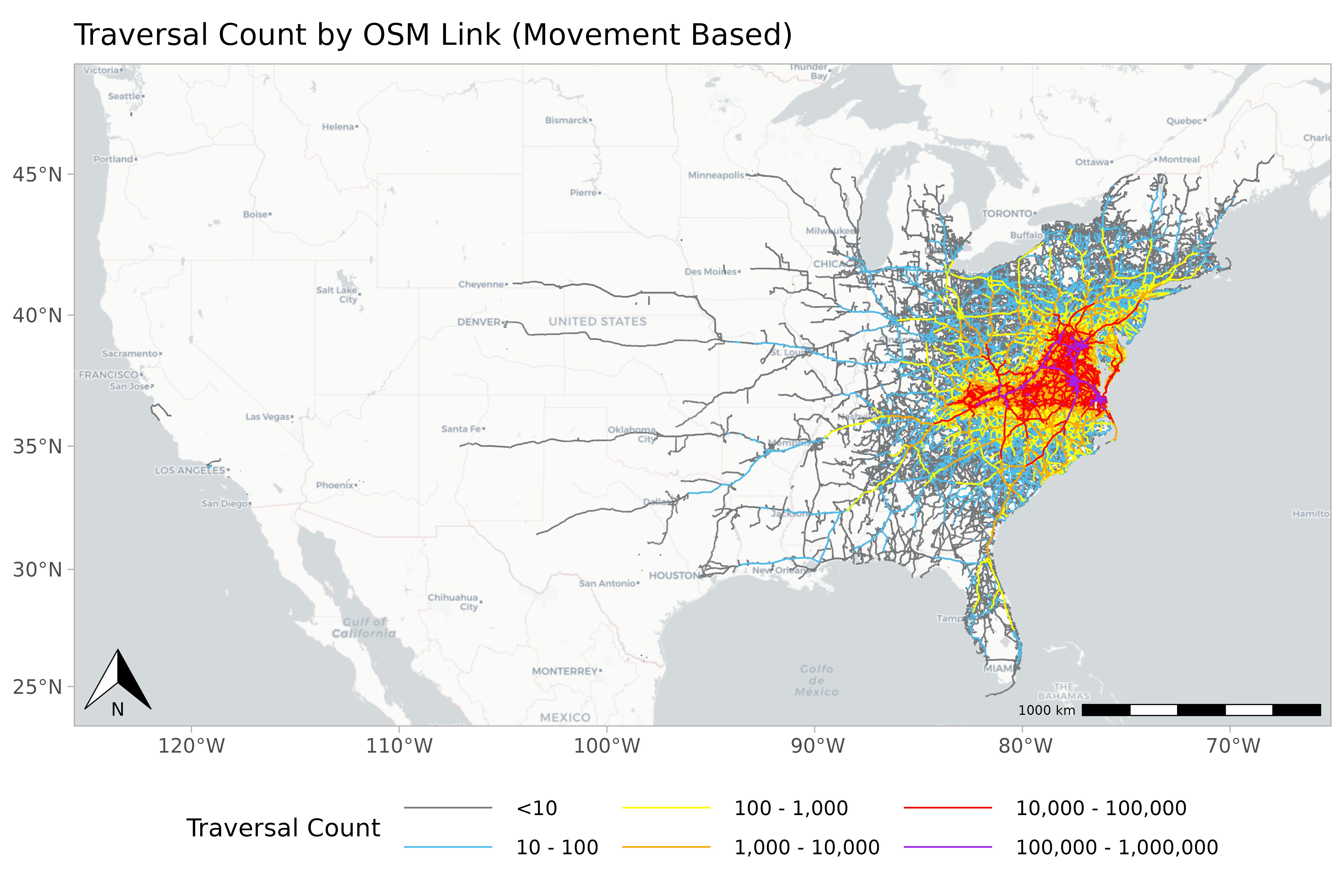}

    \centering
    \includegraphics[width=1\columnwidth,keepaspectratio]{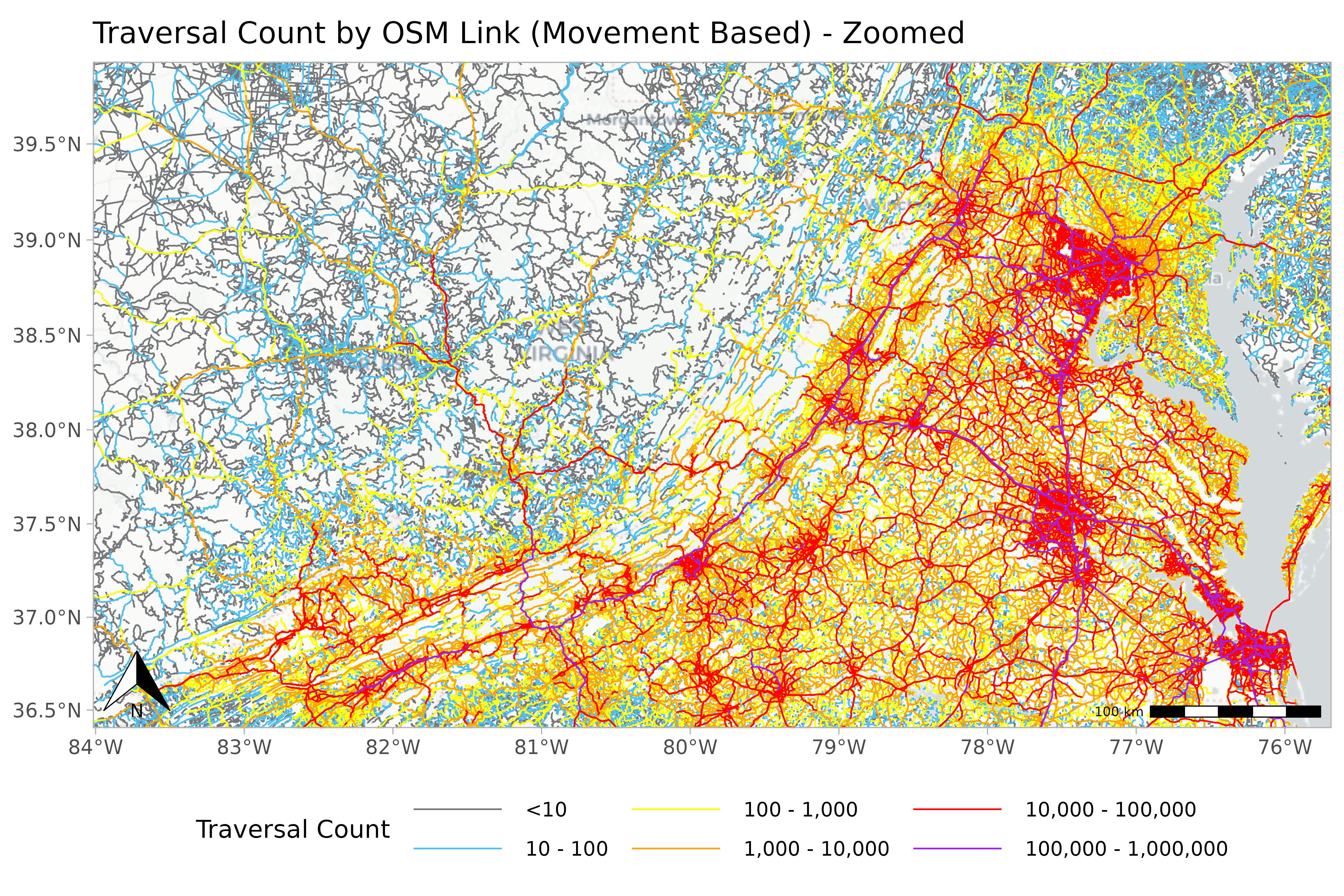}
\caption{Heatmap of traversals in the dataset grouped by roadway segment and zoomed into Virginia.}
\label{fig:traversal-heatmap-zoom}
\end{figure}

Figure \ref{fig:traversal-heatmap-zoom} illustrates the roadway geometries on which the data was collected, and the color indicates the number of traversals on each roadway segment. The geometry was acquired from OpenStreetMap, and the data were aggregated through map-matching, which is  described in the next section. The bottom part of Figure \ref{fig:traversal-heatmap-zoom} is a zoomed-in version of the same illustration, centered on Virginia. It can be seen that most roadway segments in Virginia had at least 100 traversals, major arterials and collectors had over 1,000 traversals, and highways over 10,000 traversals, with some interstates having over 100,000 traversals on each roadway segment. This shows that there is enough data on each segment for making representative conclusions. The extent of roadway segments away from Virginia shown in the top part of  Figure \ref{fig:traversal-heatmap-zoom} also illustrates that this dataset contains some fairly long trips.


\section{Methodology}
The methodology section is divided into two subsections: General Data Processing, covering the data processing tasks that were needed for all data irrespective of the eventual research questions being asked, and Insight-specific Data Processing, where the tasks are specific to the research question.

\subsection{General Data Processing}
The original data consisted of about 2 TB of parquet files split into the movement and event datasets. During initial exploration and data processing, several inefficiencies were discovered in how the data was structured. First, one trip of movement data was sometimes split across multiple parquet files. It is important to load one complete trip into memory for map-matching and trip summarization, and having that data across several parquet files was quite inefficient. Second, the data contained nested structures that were not compatible with some Python and R libraries. Therefore, the data needed to be rewritten in a more efficient format before large-scale data processing could be performed.

\subsubsection{Data Rewriting}

Before the data could be rewritten in an efficient format, it first needed to be summarized so that the file location of each trip was known. This was necessary to deterministically read a certain subset of data at a time. Therefore, a summary dataset was first created that contained the unique \textit{journey\_ids} in each parquet file. This summary dataset was then used to ensure that at least 10,000  complete trips were loaded into memory, sorted by timestamp, and written into a parquet file. The incompatible nested structures were also flattened into multiple columns during this step. Other checks such as removal of invalid or duplicate timestamps were also performed. 

The resulting dataset contained complete trips in each parquet file, which enabled reading whole parquet files into memory and processing one trip at a time without having to query the parquet dataset. This task was resource intensive and was performed on a Dask cluster with 1,536 cores and 8 TB of memory. However, if the original dataset did not have the inefficiencies, such computational resources would not be required. 

\begin{figure*}[t]
    \centering
    \includegraphics[width=0.7\textwidth,keepaspectratio]{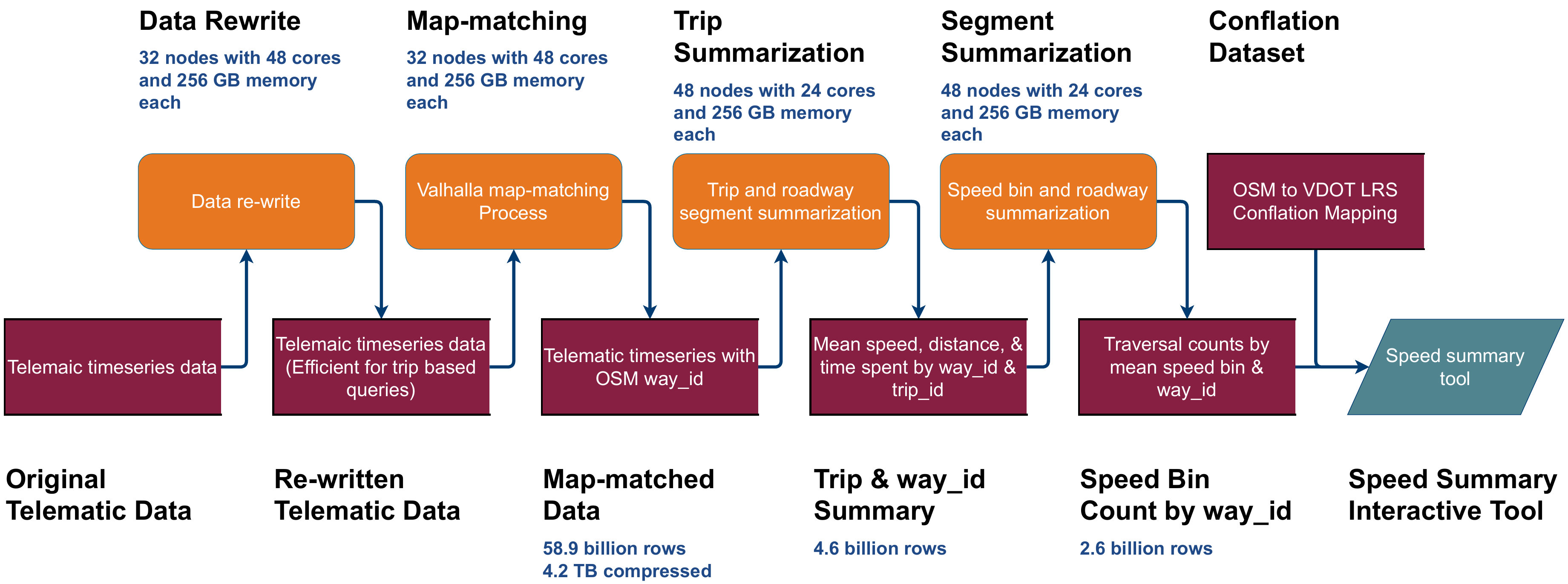}
\caption{Data flow diagram showing the various datasets being processed and combined to extract insights about roadway segments, trips, and events.}
\label{fig:data-flow}
\end{figure*}

\subsubsection{Map-Matching}
Map-matching is a process of linking vehicle trajectory data with the corresponding roadway attributes on which the vehicle trajectory was generated. This study used Valhalla, which is an open-source routing and map-matching engine that utilizes OpenStreetMap (OSM) as the basemap. Twelve instances of Valhalla were set up with two CPU threads and 16 GB of memory each. The efficient rewritten dataset was map-matched by creating multiple jobs on a parallel compute cluster, with each job sending one trip at a time to the Valhalla instances till all the trips in the dataset were processed.

For each timestamp within a trip that had valid location coordinates, the Valhalla map-matching algorithm returned a corresponding location interpolated along the centerline of the roadway geometry and a unique identifier for the roadway segment called \textit{way\_id}. Each \textit{way\_id}, in turn, was associated with OSM parameters such as roadway classification, speed limit, and number of lanes. In addition to the attributes normally available through OSM, each roadway segment was also conflated with other attributes available through the VDOT Linear Referencing System (LRS) map \cite{Ali2025ITSCLRSConflation}. The resulting dataset contained all the columns of the efficient rewritten dataset, as well as the \textit{way\_id} for each data point representing the roadway segment on which it was collected. 

Figure \ref{fig:data-flow} illustrates the above-mentioned processing tasks as the first two orange blocks of the data processing pipeline. The first three maroon blocks represent the original telematics data, efficient rewritten telematics data, and the map-matched data. The map-matched dataset consisted of 58.9 billion rows and occupied 4.2 TB of hard drive space. This dataset could now be used for a variety of different analysis.

\subsection{Insight-specific Data Processing}
This paper focuses on two use cases: developing a speed profile for every roadway segment in Virginia and understanding the trip-taking behavior of drivers by various zip codes. Development of speed profiles for roadway segments would support a number of traffic engineering applications, such as speed limit studies and safety studies that look at crash causes and countermeasures.  Data on the distribution of trips between zip codes provides valuable information that can be used for the development of travel demand models that are used to forecast travel demands and plan for where new transportation infrastructure may be needed.  The following subsections discuss the data processing steps for each use case.

\subsubsection{Roadway-level Summarization}
As shown in Figure \ref{fig:data-flow}, the roadway-level summarization involved two additional steps after the general data processing was completed. First, a trip- and \textit{way\_id}-level summary was created to generate metrics such as mean, maximum, minimum, and standard deviation of speed, total time spent on \textit{way\_id}, number of data points on \textit{way\_id}, date, hour of day, etc. All of these metrics correspond to one continuous traversal of one vehicle on the particular roadway segment represented by the \textit{way\_id}. This dataset consisted of 4.6 billion rows once the data processing was completed.

Second, a \textit{way\_id}-level summary was created by counting the number of traversals in various mean speed bins by hour of day and date. Therefore, for any given date and hour, the speed profile of any roadway segment can be summarized by the number of traversals in each speed bin. 

\subsubsection{Trip-level Summarization}

The trip-level summary consists of metrics such as trip start and end location, trip start and end zip codes, duration, trip length, distance between start and end locations, and number of data points. Such summaries can be used to understand origin and destination relationships between various regions of the state at different times of day. Metrics were extracted using precise locations as well as zip codes, so that the right balance could  be achieved between accuracy of metrics and protection of personally identifiable information.

The trip-level summaries were created using both the movement data and the event data. However, since the overlap between the two datasets was quite small and the movement dataset provided more granular information, only the movement-based trip summaries are discussed in the results section.

\section{Results}

\subsection{Roadway Speed Summaries}

\begin{figure*}[!p]
    \centering
    \includegraphics[width=0.9\textwidth,keepaspectratio]{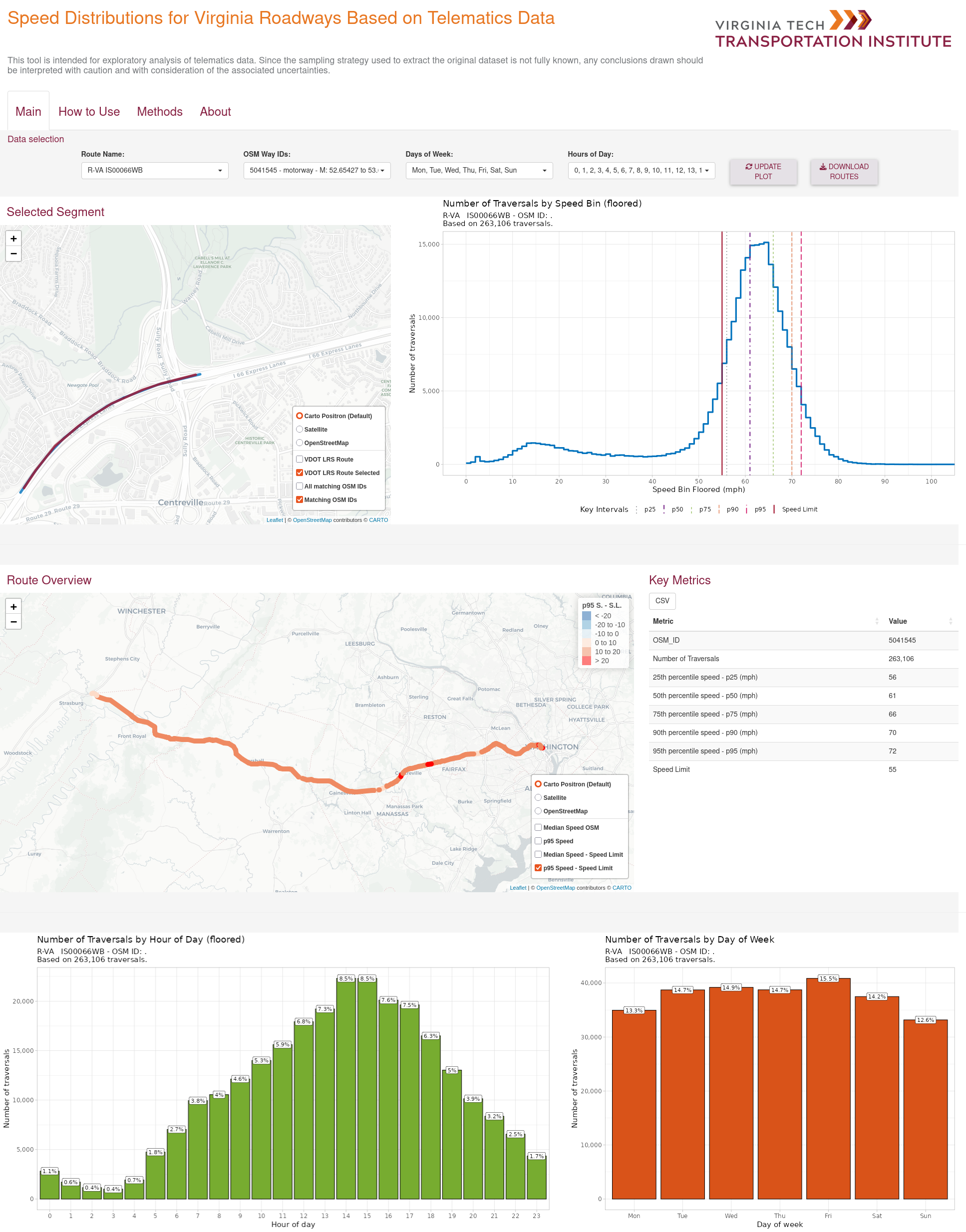}
\caption{Screenshot of speed distribution tool visualizing various aspects of data such as histograms, route heatmaps, data distribution, and key metrics (\href{https://dataviz.vtti.vt.edu/roadway_speed_summary_app/}{accessible online}).}
\label{fig:roadway-speed-sum}
\end{figure*}

\begin{figure*}[t]
    \centering
    \includegraphics[width=0.7\textwidth,keepaspectratio]{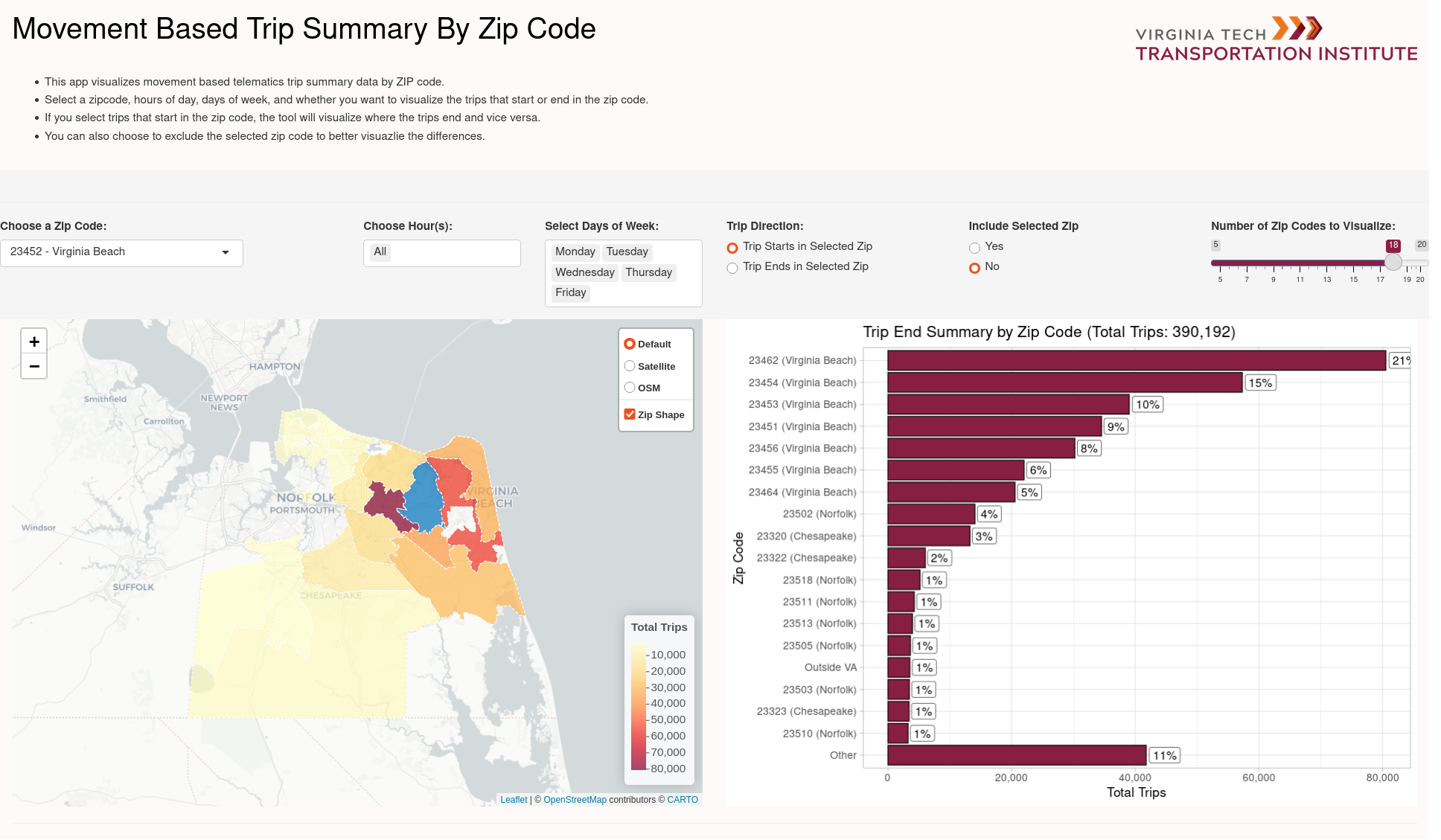}
\caption{Screenshot of trip summary visualization tool displaying distribution of destinations for selected zip code (\href{https://dataviz.vtti.vt.edu/Movement-Trip-Summary-By-Zip-Code/}{accessible online}).}
\label{fig:OD-analysis}
\end{figure*}

Figure \ref{fig:roadway-speed-sum} shows a screenshot of the speed distribution tool, \href{https://dataviz.vtti.vt.edu/roadway_speed_summary_app/}{accessible online}. This tool illustrates the various types of metrics that have been extracted and can be visualized for each roadway segment in Virginia. 

The data selection inputs enable the user to select a route name, a roadway segment represented through a \textit{way\_id} and mile markers, days of week, and hours of day that they want to analyze. In the illustrative example, route I-66 westbound and the roadway segment representing mile marker 52.6 to 53.6 is selected for all days of week and hours of day.

The first map in the top row of visualizations displays the geometries of the roadway segment according to the VDOT linear reference system (LRS) and OSM maps. The second visualization in the top row shows a histogram representing the number of traversals on that roadway segment in each speed bin. There are 263,106 traversals for this selection and the distribution is bimodal, with the majority of the traversals happening between 55 and 75 MPH. A smaller secondary mode can also be seen between 10 and 20 MPH.  The particular section of road often experiences congestion in afternoon peak periods, which is the cause of this secondary low speed mode. 
 This type of data can be useful for DOTs for better understanding traffic congestion patterns  for various roadway segments. Key metrics like 25\textsuperscript{th}-to 95\textsuperscript{th}-percentile speeds and speed limits are visualized on the plot and also provided as a table in the second row.

The map in the second row visualizes all the roadway segments for the selected route using median speed, 95\textsuperscript{th}-percentile speed, difference between median speed and speed limit, and difference between 95\textsuperscript{th}-percentile speed and speed limit. This map gives an overview of the route and helps users identify segments that are outliers on any of these metrics. 

The third row visualizes the distribution of traversals by hour of day and day of week to help the user understand how much data was accumulated in each of these bins. The user can iteratively modify a selection to refine their analysis and better understand contributing factors. This tool can be used to visualize the speeding profile of any major roadway segment in Virginia and quantify driving behavior on it.

\subsection{Zip-level Origin and Destination Analysis}
Figure \ref{fig:OD-analysis} illustrates the trip summary interactive visualization tool that enables users to analyze zip-level origin and destination data for Virginia. This tool is also openly \href{https://dataviz.vtti.vt.edu/Movement-Trip-Summary-By-Zip-Code/}{accessible online}.
The top row shows various data selection options where the user can choose a zip code within Virginia, hours of day, days of week, whether the zip code should be treated as origin or destination, and whether the trips that start and end within the zip code should be considered in the summary. The analysis in the illustrative example shows data for trips starting in zip code 23452 and ending outside it. 

The first plot in the second row shows a heat map of destinations for trips originating in the selected zip code. The second plot displays the numbers of trips and relative percentages through a bar chart. This informs the general traffic flow trends for a given zip code at a given hour of day.

\begin{figure}
    \centering
        \includegraphics[width=0.95\linewidth]{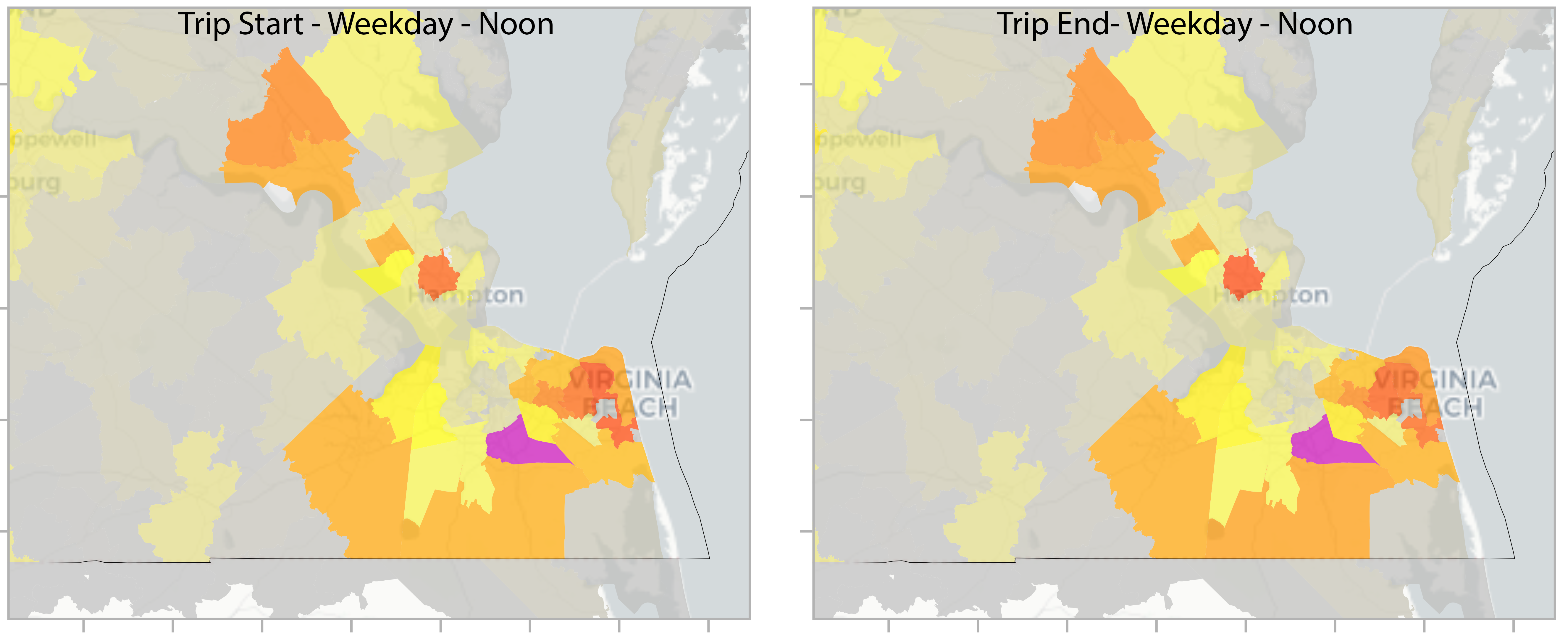}  
        \includegraphics[width=0.95\linewidth]{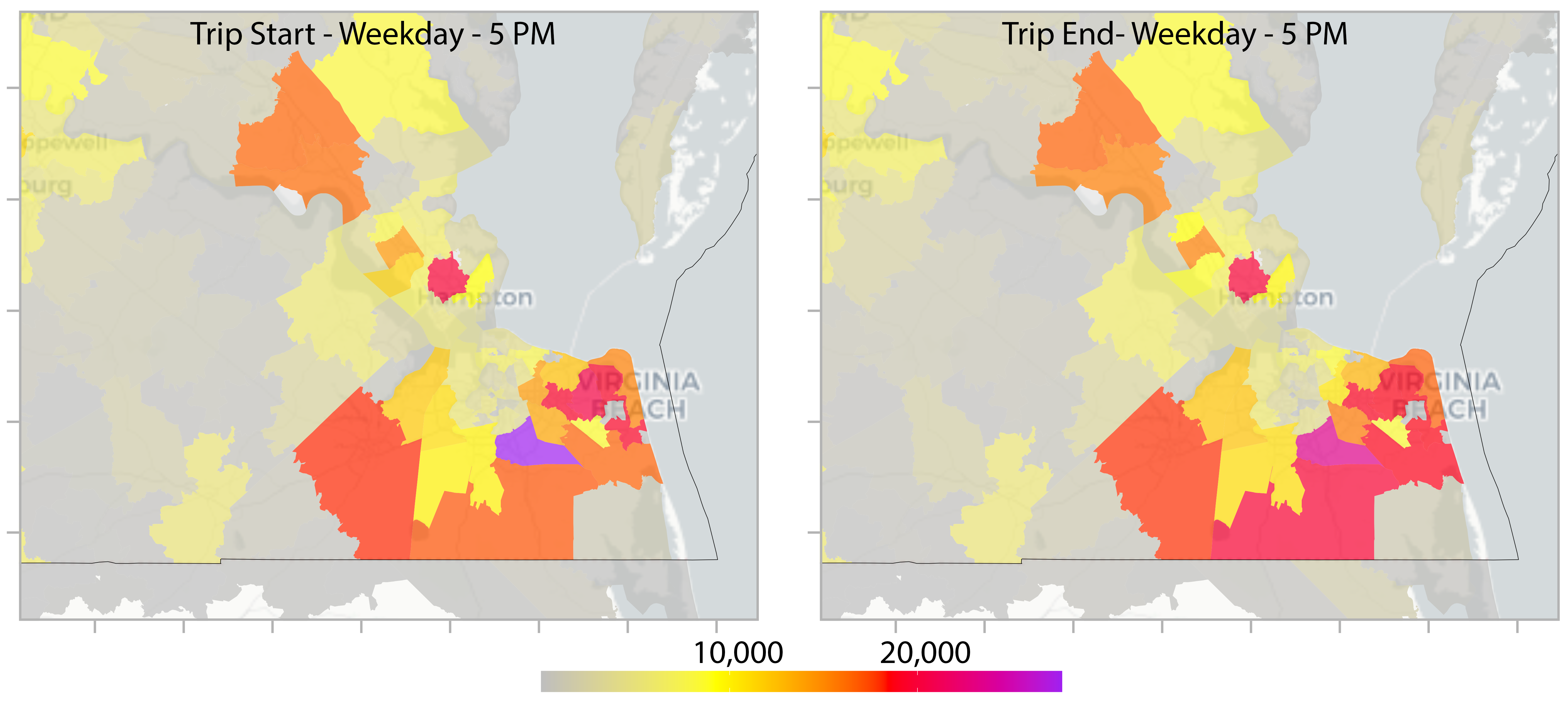}  
        \hfill
        \noindent\color{gray}\rule{0.95\linewidth}{1pt} 
        \hfill
        \includegraphics[width=0.95\linewidth]{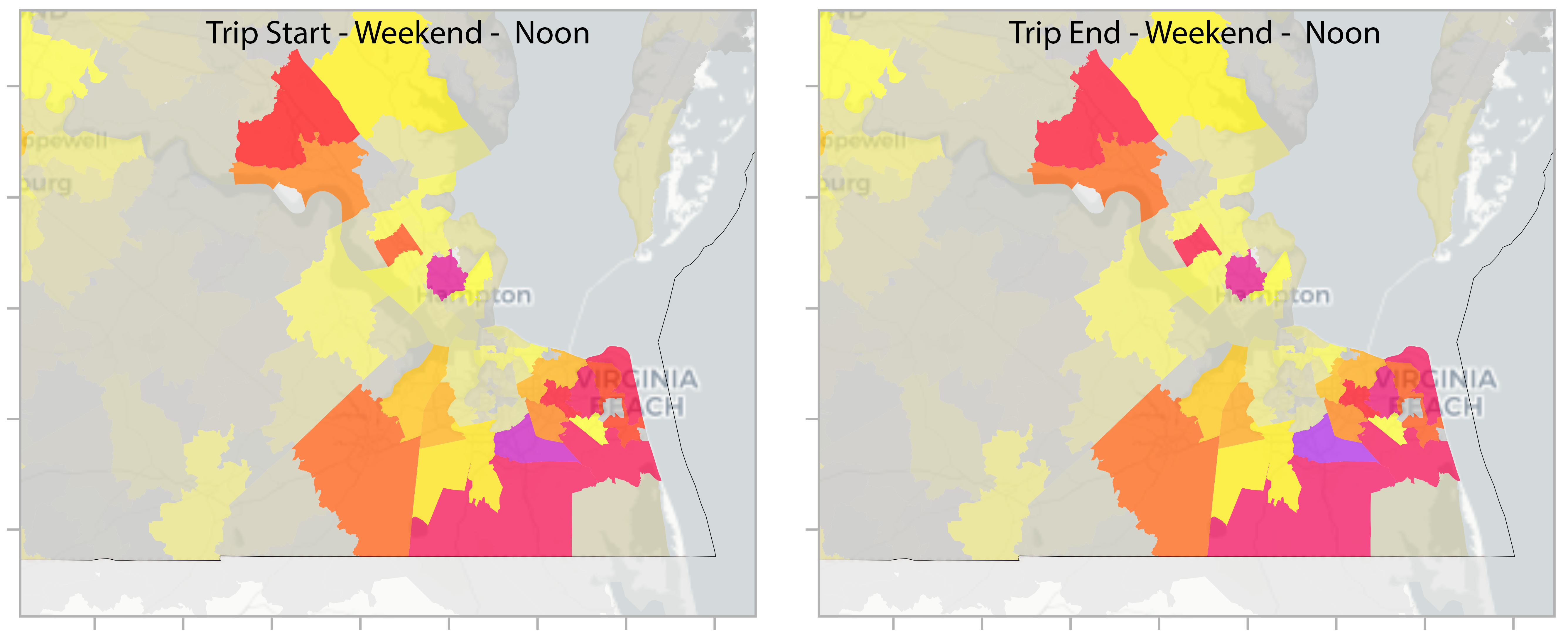}  
        \includegraphics[width=0.95\linewidth]{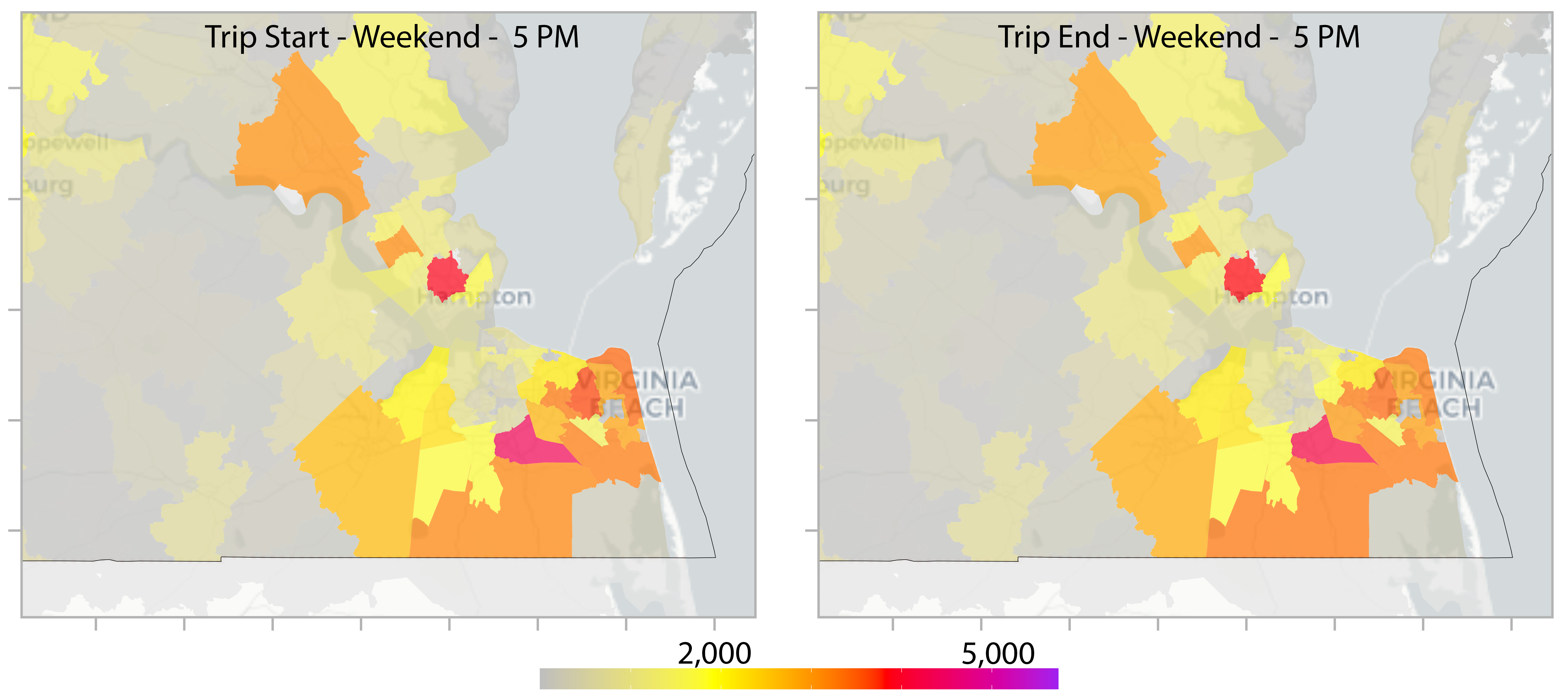}  
    \caption{Comparison of trip start and end heatmaps for noon and 5 p.m. on weekdays versus weekends around Hampton Roads region.}
    \label{fig:heatmap-comp}
\end{figure}

Figure \ref{fig:heatmap-comp} shows another way of analyzing zip-level origin and destination data. The figure shows heat maps of trip start and ends for the Hampton Roads region of Virginia at 12 p.m. (noon) and 5 p.m. with the data for the weekdays at the top and data for weekends at the bottom. The figure shows that from a perspective of trip percentages, the 5 p.m. trends for the weekdays are closer to the 12 p.m. trends for weekends. 

When comparing the weekday trip start versus trip end heatmaps at 5 p.m. (second row in Figure \ref{fig:heatmap-comp}), it can be clearly seen that a lot more trips are starting in the Chesapeake area (23320), shown in purple, and ending in regions surrounding it. This is another illustrative example of how telematics summaries can inform planners and researchers about traffic activity for the whole state from a single data source and processing pipeline. Such analyses can also be performed using other spatial aggregation units such as census tracts, geohashes, and traffic analysis zones, depending on what aggregators are used by the DOT.

\section{Discussion}

This paper shows how large-scale telematics data can be processed to extract insights for DOT applications. Such data can provide a wealth of information about driving behavior and traffic flow that is useful for traffic planners and researchers.  The key takeaways and recommendations from this work can be summarized up as:

\begin{itemize}

    \item These data can be highly effective for a number of DOT applications, such as understanding speed-related driving behavior and traffic trends at scale. 
    
    \item Telematics data has several key advantages. First, the scale of the data and the manufacturer-level availability make it more useful and generalizable than other similar data sources collected on a smaller scale.  Additionally, since the data is sourced from manufacturer-installed sensors, the probability of failures and standardization errors is smaller than other data sources. Second, the continuous nature of the data collection makes the data available for all roadways as compared to data collection methods that require hardware installation on infrastructure  elements. Finally, the single source access and single pipeline processing make the insights more dependable and sources of error much easier to trace.

    \item To extract meaningful insights, it is important to understand the underlying sampling biases of the data. Therefore, it would be helpful to have metrics about vehicle classification and intended usage, such as personal versus commercial use. 

    \item Even though the data architecture and processing pipeline will be driven by the applications utilizing the data, some general data processing steps should be introduced that will benefit many potential applications. For example, being able to efficiently analyze full trips is important for map-matching and summarization. Therefore, if the data applications require map-matching or trip summarization, OEMs and data aggregators should package full trips in one file. This will reduce the computational needs of processing the data.

    It is important to understand how the various steps in the data pipeline would differ as the data scales or the requirements of the tool change. The first three data stores in the data pipeline \--- the original telematics data, rewritten telematic data, and map-matched telematic data \--- would all scale linearly as the number of trips increased. However, the data resources needed for these transformations could be less intensive if the data were stored and processed as complete trips when  collected from  vehicles. 

    The trip- and \textit{way\_id}-level summaries would also scale linearly as the number of trips increases. However, the speed bin count by \textit{way\_id}  does not increase with the number of trips but depends on the number of roadway segments. If there is a need to filter the data by date, the size of this dataset would also depend on the date range.
    
    The speed summary interactive tool displays the speed distribution of one roadway segment at a time. If the combined speed distribution of multiple roadway segments is required \--- such as for a custom mile marker range for a given route \--- the tool would need to interface with the trip and \textit{way\_id} summary table instead of the speed bin count table. This would  have consequences for data processing latency and intermediate table sizes. Therefore, there is a need for better understanding the major use cases of this data and standardization of data processing pipelines across manufacturers to support the DOT needs.   
    
\end{itemize}

\section{Conclusions and Future Work}

Within the next few years, a majority of vehicles on US roadways will be sharing data about their location, kinematics, and vehicle states at regular intervals through telematics data. This presents a transformational opportunity for intelligent vehicle systems. However, the many challenges presented by the scale of the data and lack of standardization  prevent us from fully leveraging this data resource.

There is a clear need for better understanding of key use cases and developing standardized practices across OEMs and data aggregators. These practices would not only support the key use cases but would also make it a lot more cost-effective from a data processing perspective. The authors will continue to explore other use cases, such as quantifying the effect of weather, analyzing the data from a safety case perspective and comparing insights derived from naturalistic driving data \cite{ali_characterizing_2023, gali_2021, ali_quantifying_2021, Beale2025ITSCConflation}.  


\section*{Acknowledgment}
The work presented in this paper was funded by the National Surface Transportation Safety Center for Excellence (NSTSCE) at Virginia Tech Transportation Institute and the Virginia Department of Transportation.

\section*{Data and Resources}
This paper is accompanied with two interactive visualization tools (\href{https://dataviz.vtti.vt.edu/roadway_speed_summary_app/}{https://dataviz.vtti.vt.edu/roadway\_speed\_summary\_app/} \& \href{https://dataviz.vtti.vt.edu/Movement-Trip-Summary-By-Zip-Code/}{https://dataviz.vtti.vt.edu/Movement-Trip-Summary-By-Zip-Code/}) and a data repository (\url{https://github.com/gibran-ali/telematics-data-analysis}). Please allow a few minutes for the interactive data visualization to load. The data available through the visualization tools and github repository are aggregated and therefore do not contain personally identifiable information (PII).

\bibliography{ITSC_References.bib}
\bibliographystyle{ieeetr}

\vspace{12pt}

\end{document}